\documentstyle[epsf,seceq]{ptptex}

\def\be{\begin{equation}}
\def\ee{\end{equation}}

\def\simm#1{\mathop{\vtop{\ialign{##\crcr
        $\hfil\displaystyle{#1}\hfil$\crcr\noalign{\kern0.5pt\nointerlineskip}
        $\sim$\crcr\noalign{\kern0.5pt}}}}\limits}

\def\P{{\sf \bf P}}
\def\R{{\sf \bf R}}
\def\W{{\sf \bf W}}
\def\C{{\sf \bf C}}
\def\PW{{\sf \bf PW}}
\def\PC{{\sf \bf PC}}
\def\RW{{\sf \bf RW}}
\def\RC{{\sf \bf RC}}

\title{%        %You can use \\ for explicit line-break
{\normalsize \hfill {\sf UTCCP-P-30}} \\
{\normalsize \hfill {\sf Nov.\ 1997}} \\ \mbox{}\\
Developments in Finite Temperature QCD on the Lattice with Dynamical Quarks%
\thanks{Review presented at the International Workshop on 
``Physics of Relativistic Heavy Ion Collisions'',
Kyoto, Japan, 9--11 June 1997. To be published in the Proceedings
[Prog.\ Theor.\ Phys., Suppl.]}
}

\author{%       %Use \sc for the family name
Kazuyuki {\sc Kanaya}
}

\inst{%         %Affiliation, neglected when [addenda] or [errata]
Center for Computational Physics and Institute of Physics, 
\\ 
University of Tsukuba, Tsukuba, Ibaraki 305, Japan
}

\recdate{%      %Editorial Office will fill in this.
\today
}

\abst{
I report on some recent developments in nonperturbative studies 
of finite temperature QCD with dynamical quarks on the lattice. 
I discuss new studies of improved lattice actions 
and their application to finite temperature QCD.
I also summarize the status of lattice investigations about the order of the 
finite temperature QCD transition for the case of two flavors of 
degenerate light quarks, using both staggered and Wilson lattice fermions.
}
\begin{document}

\maketitle

\section{Introduction}

The finite temperature transition in QCD is characterized by 
recovery of the spontaneously broken chiral symmetry 
and also by breakdown of the confinement 
at temperatures higher than a critical value $T_c$.
Because both of these characteristic properties of QCD are essentially 
non-perturbative in nature, the most systematic way of studying 
the transition is to investigate it on the lattice. 

A large number of 
numerical simulations have been performed to clarify
the nature of the QCD transition from the first principles of QCD.
In order to make a precise prediction for the real world 
from the results obtained on finite lattices, 
we perform an extrapolation to the limit of vanishing 
lattice spacing, the continuum limit. 
For a reliable extrapolation, 
we have to simulate on sufficiently fine lattices. 
On the other hand, in a study of finite temperature QCD,
the spatial lattice size should be sufficiently large in order
to approximately realize the thermodynamic limit.
With the limitation of the currently available computer power, 
this means that 
we have to suppress the lattice size in the time direction $N_t$; 
currently $N_t \approx 4$--6 is used for major simulations of
QCD with dynamical quarks (full QCD). 
Because the temperature on the lattice is given by $T=1/N_t a$ with $a$ 
the lattice spacing, the corresponding lattices become rather coarse; 
for $T \sim T_c \approx 100$--200MeV, $a \sim 0.2$--0.4fm.
Subsequent lattice artifacts sometimes make the analysis and 
interpretation of lattice results not straightforward.

Here, recent developments in improved lattice actions opened us a
possibility to reduce the lattice artifacts already on coarse lattices 
and, therefore, to extract accurate results for finite temperature QCD 
transition with the present computer power.
In section~\ref{sect:impr}, I report these developments in improved lattice 
actions and their application to finite temperature QCD.

In a phenomenological study of quark-gluon plasma in heavy ion collisions 
and in the early Universe, the most fundamental information required
is the order of the QCD phase transition.
Recently, there have been several developments in lattice QCD on this issue, 
especially for the case of two flavors of degenerate light quarks. 
In sect.~\ref{sect:scaling}, I report about
the status of these recent studies 
using both staggered and Wilson lattice fermions.
This also includes one of the first fruitful applications of 
improved actions to the study of QCD at finite temperatures.

For other developments in lattice QCD, I refer reviews in recent Lattice 
conferences\cite{ref:LatticeXX}.

\section{Improvement of the lattice action}
\label{sect:impr}

An action on the lattice is chosen such that the continuum 
action is recovered when we let the lattice spacing $a \rightarrow 0$ in the 
action. 
For the gluon part of the action, the minimal choice is the standard 
one-plaquette action by Wilson\cite{wilson74}, which is the most local 
and geometrically the simplest lattice gauge action.
On the lattice, however, we have freedom to introduce less local 
terms to the action without affecting the continuum limit. 
Here, the speed of approach to the continuum limit does depend 
on the choice of the action. Therefore, a judicious choice of the action 
can suppress lattice artifacts in physical observables even at 
moderate values of the lattice spacing. 
Such actions are called ``improved actions''. 
Importance of improvement is much stressed recently
(for reviews see Refs.~\cite{IMreviewH,IMreviewL}).

The basic idea behind improvement may be obtained by considering the 
lattice derivative;
the naive derivative 
$\Delta_x f(x) = {1 \over {2a}} [f(x+a)-f(x-a)]$
converges to the continuum derivative with an error of $O(a^2)$.
We can reduce this error down to $O(a^4)$ by replacing 
$\Delta_x \rightarrow \Delta_x - {a^2 \over 6}\Delta_x^3$ 
which operates on fields at up to next nearest neighbor points.
Similar substitution is effective also to obtain a lattice action
that approaches to the continuum action much smoothly.
The life is, however, not so easy, because what we want to obtain
is not a smoother lattice action, but an action which leads to smaller
discretization errors in the physical observables 
$\langle P \rangle = Z^{-1} \int {\cal D}\phi P[\phi] e^{-S[\phi]}$
which can be obtained only after a non-perturbative calculation, 
in general.
Many different strategies have been proposed to obtain such $S$.
Two major approaches are the renormalization group methods proposed
first by Wilson, and the perturbative improvements by Symanzik.
In any case, improved lattice actions contain additional terms 
(interactions that usually have a wider spatial size) 
compared with the standard action. 
Because such additional terms 
makes the simulation quickly difficult and time-consuming, 
the efficiency of improvement should be tested for each of the new 
terms introduced.

\subsection{Comparison of improved actions}
\label{sec:CPPACS}

\begin{figure}[t]
\centerline{
\epsfxsize=12cm \epsfbox{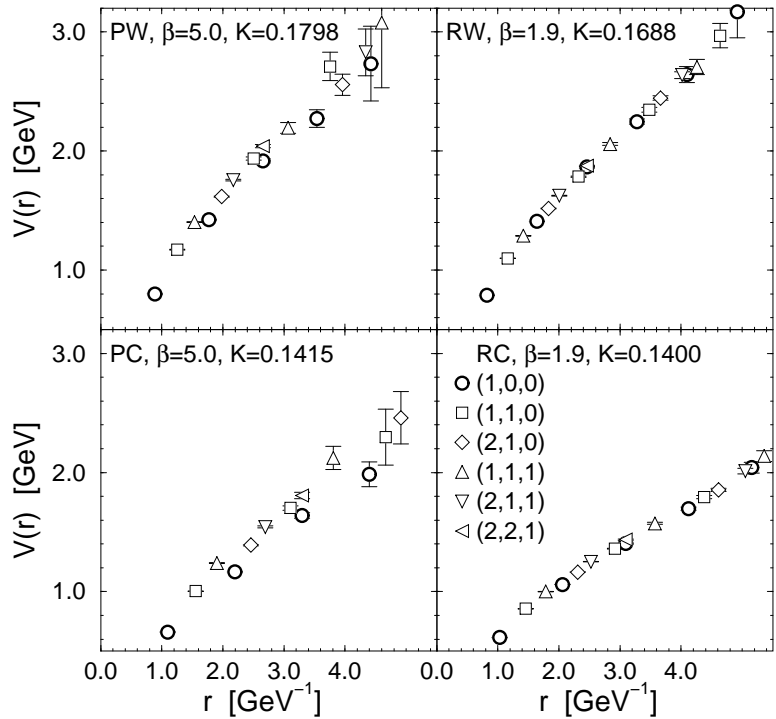}
}
\vspace{-0.2cm}
\caption{
Static quark potential obtained by 
a full QCD simulation %for two-flavor QCD 
performed on $12^3\times32$ lattices 
using various lattice actions.\protect\cite{CPPACSf}
The coupling parameters $\beta=6/g^2$ and $K$ are adjusted 
such that we obtain the lattice spacing $a \approx 0.2$fm and 
the ratio of the pseudo-scalar and vector meson masses 
$m_{\rm PS}/m_{\rm V} \approx 0.8$.
}
\label{fig:potential}
\end{figure}

\begin{figure}[t]
\vspace{-0.3cm}
\centerline{
\epsfxsize=9cm \epsfbox{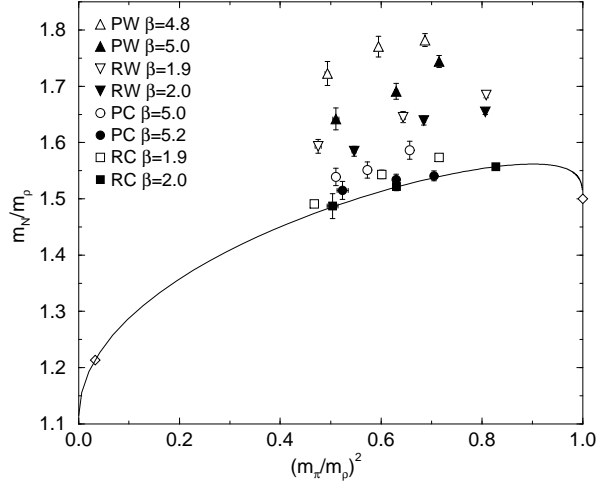}
}
\vspace{-0.5cm}
\caption{
Octet baryon mass as a function of PS meson mass squared,
normalized by vector meson mass at $a \approx 0.2$fm 
using various lattice actions.\protect\cite{CPPACSf}
The diamond near the lower 
left corner is the experimental point computed with N, $\pi$ and $\rho$.
}
\label{fig:spectrum}
\end{figure}

Improved lattice actions have been studied mainly in quenched QCD 
(QCD without dynamical quark loops). %\cite{quenchedIMtest}. 
In finite temperature QCD, however, dynamical quarks are essential.
Quite recently, the CP-PACS Collaboration have performed a systematic 
investigation of how various terms added to the gauge and quark actions, 
taken separately, affect light hadron observables in full QCD\cite{CPPACSf}.
A new dedicated parallel computer CP-PACS\cite{cppacs} is used 
for the simulation.
The effect of improvement 
is studied comparing,
for the gauge part of the action, the standard one plaquette action (\P) 
and an improved action (\R) that 
contain $1\times2$ rectangular loops with the coefficients obtained 
by a renormalization group study\cite{Iwasaki83},
and for the quark action, the standard Wilson quark action (\W) 
and an improved Wilson quark action containing 
the so called clover term\cite{clover} (the clover action: \C).

Simulations are carried out in two-flavor QCD. 
Because the extent of improvement to be clearer at a coarser lattice
spacing, we adjust the gauge coupling parameter $6/g^2$ 
such that the lattice spacing becomes $a \sim 0.2$fm. 

Figure \ref{fig:potential} summarizes the results for the static quark 
potential using the four different combinations for the 
action (\PW, \PC, \RW, and \RC).
Different symbols correspond to potential data measured in different 
spatial directions on the hypercubic lattice, 
along the vector indicated in the figure.
The jagged result for the standard action (\PW) means that the
rotational symmetry is sizably violated on this coarse lattice.
A drastic improvement of the rotational symmetry is seen 
when we replace the gauge action with the \R\ action, 
as observed previously also in the quenched case\cite{ourPot97}.
In this regard, the quark action has much less effect.
Fig.~\ref{fig:spectrum} shows the results for light hadron masses. 
The solid curve represents the result of 
a phenomenological mass formula by Ono\cite{ONO},
which is approximately reproduced also on the lattice 
when $a$ is sufficiently small 
($a \simm{<} 0.1$fm for the \PW\ action).
We see that improvement of the quark action makes the spectrum quite 
close to the phenomenological curve already at $a \approx 0.2$fm.

These results are very encouraging: 
They show that, even with the presently available computing power, 
it may be possible to obtain accurate results near the continuum
limit also in full QCD, when we use an improved action.

\subsection{Improved actions for full QCD at finite temperatures}
\label{sec:ImFT}

One of the first systematic applications of improved action 
to the study of finite temperature QCD has been done by 
the Tsukuba group\cite{ImFQcdTsukuba,ourPRL97}.
Their motivation to adopt an improved action is 
to remove severe lattice artifacts encountered in a finite temperature
simulation of QCD with the standard Wilson quark 
action\cite{MILC46,ourStandard96}. 
Using the \RW\ action discussed in the previous subsection, these 
lattice artifacts are shown to be well removed.
Further physical outputs from this study will be discussed 
in Sec.~\ref{sec:QCDPAX}.

Through this study, it became clear that improvement of the lattice action
is essential in order to study several important topics in finite 
temperature QCD\cite{KanayaLat95}. 
Other major research groups have also began to study 
the finite temperature QCD with improved actions\cite{UkawaLat96}. 

The MILC Collaboration combined a Symanzik type improved gauge action 
and the clover quark action \C, and found good improvement\cite{ImFQcdMILC_FT} 
similar to that observed in the case of the \RW\ action.
Comparing these results together with an unpublished \PC\ data\cite{AUU96}, 
where the lattice artifacts are found to be remaining, 
we conclude that the big reduction of the lattice artifacts
noted in the simulation of finite temperature QCD is 
mainly due to improvement of the gauge action.
The MILC Collaboration studied also an improved staggered quark action, 
aiming at a better flavor symmetry on a coarse lattice.

The Bielefeld group also extended their study of improved pure gauge 
theories\cite{ImQQcdBielefeld} 
to include improved staggered and Wilson quark actions\cite{ImFQcdBielefeld}.
One of their motivations is to achieve a faster convergence 
of thermodynamic quantities to the large $N_t$ limit:
Thermodynamic quantities have errors of $O(1/N_t^2)$ on the lattice 
because the relevant fluctuations include 
those with spatial size $\sim 1/T = N_ta$, 
that feel discretization errors when $N_t$ is not large enough. 
These errors are automatically removed when we take the continuum limit 
$a \rightarrow 0$
keeping the temperature $T=1/N_ta$ fixed, i.e., $N_t \rightarrow \infty$
simultaneously. 
[Practically, $N_t \simm{>} 8$ is required to obtain thermodynamic 
quantities at $T \sim O(T_c)$ to an accuracy of several percent 
in quenched QCD using the standard one-plaquette action.]
Because an improved action has improved short-distance properties, 
we expect these errors to be also smaller with improved actions.%
\footnote{
The Bielefeld group extended the criterion to include improvement 
also in the ``high temperature limit'' on lattices 
with a fixed small $N_t \sim 4$: 
Since $a$ is a decreasing function of $6/g^2$, a larger $6/g^2$ 
corresponds to a larger $T=1/N_ta$.
For fixed $N_t$, however, the ``high temperature limit'' 
$6/g^2 \rightarrow \infty$ corresponds to a temperature higher than
$T=\infty$ in the usual procedure in which the limit $T \rightarrow \infty$
is taken only after the limit $a \rightarrow 0$ with fixed $T$.
Because this includes an exchange of the limiting procedures, 
I think this extended criterion to be slightly different in nature 
from the usual criterions for improvement concerning the convergence 
to the continuum limit. 
The difference becomes important when the number of 
adjustable coupling parameters in the improved action is restricted.
}

In spin models and also in quenched QCD, good improvements are reported
using a class of improved actions called
fixed point actions (``perfect actions''),
which are designed to be at the UV fixed point of a renormalization group
transformation and therefore to remove all lattice artifacts when 
infinite number of coupling parameters are introduced\cite{IMreviewH}.
Trials to construct a fixed point action for full QCD are also 
made\cite{IMreviewH,Wiese97,DeGrand97}.

\section{Order of the chiral transition in two-flavor QCD}
\label{sect:scaling}

Let us now turn our attention to the topics of the order of 
of the QCD transition from the high temperature
quark-gluon-plasma phase to the low temperature hadron phase,
which is supposed to occur at the early stage of 
the Universe and possibly at heavy ion collisions.
It is, in particular, crucial to know 
whether the transition is a first order phase transition
or a smooth transition (second order phase transition or crossover).

The nature of the QCD transition is considered to 
depend on the number of quark flavors $N_F$ sensitively.
The physically interesting cases are the case 
of two and three flavors of degenerate light quarks ($N_F=2$ and 3), 
and also a more realistic case of two light and one heavy quarks ($N_F=2+1$). 
The case $N_F=2$ corresponds to the case where the third 
quark s is much heavier than the relevant energy scale
for thermal processes near the critical temperature,
$m_s \gg T_c$, 
while the case $N_F \geq 3$ corresponds to the case $m_s \ll T_c$.

Using universality hypothesis, 
the nature of the finite temperature QCD 
transition near the chiral limit %(the limit of vanishing quark masses) 
can be studied by a Ginzburg-Landau effective theory
respecting the chiral symmetry of QCD, 
the effective $\sigma$ model\cite{PisarskiWilczek}.
For $N_F \geq 3$, a first order transition is predicted.
For $N_F=2$, on the other hand, the order of the transition 
is not quite definite in the effective $\sigma$ model; 
a first order transition is predicted when the anomalous axial 
U$\!_A$(1) symmetry 
is effectively restored at the transition temperature, 
while a second order transition is expected otherwise.
Because the U$\!_A$(1) breaking 
operator is a relevant operator whose coefficient grows 
towards the IR limit under a renormalization group transformation, 
the transition is more likely to be second order \cite{Ukawa95}. 
A non-perturbative study is required to determine the order of the 
transition conclusively. 

Therefore, understanding the nature of the QCD transition for $N_F=2$ is 
an important step toward the clarification of the transition 
in the real world. 

When the chiral transition is second order, we expect that 
the transition turns into an analytic crossover at non-zero $m_q$, 
while when the chiral transition is first order, it will remain
to be first order for small $m_q$. 
However, it is difficult to numerically distinguish between 
a very weak first order transition and a crossover, 
especially at small $m_q$:
In order to confirm the expected crossover numerically, 
we have to study the lattice size dependence to see if the 
formation of singularity 
(e.g.\ the increase of the peak height of a susceptibility with increasing 
the lattice volume) stops on sufficiently large lattices.
This becomes more and more difficult at small $m_q$.

Here, the universality provides us with useful scaling relations 
that can be confronted with numerical results of QCD,
in order to test the nature of the transition: 
It is plausible from an effective $\sigma$ model
that, when the chiral transition %(transition in the chiral limit)
is of second order, QCD with two flavors
belongs to the same universality class as the three 
dimensional O(4) Heisenberg model\cite{PisarskiWilczek}.
The O(4) model is much simpler than the $\sigma$ model,
and its scaling properties are well studied. 
For example, 
at small external field $h$ near the critical temperature $T_c$ for $h=0$, 
the pseudo-critical temperature $T_{pc}(h)$ and 
the peak height of the magnetic and thermal susceptibilies 
follow $T_{pc}-T_c \sim h^{z_g}$, 
$\chi_m^{\rm max} \sim h^{-z_m}$, and 
$\chi_t^{\rm max} \sim h^{-z_t}$, 
where $z_g = 1/\beta\delta$, $z_m=1-1/\delta$, 
and $z_t=(1-\beta)/\beta\delta$
in terms of the O(4) critical exponents $\beta$ and $\delta$. 
Here the values of $\beta$ and $\delta$ for the O(4) model 
are well established\cite{KanayaKaya}. 
In QCD, we identify $T \sim 6/g^2$, $h \sim m_q$, and 
$M \sim \langle \bar\Psi \Psi\rangle$. % the chiral condensate. 

In lattice QCD, an additional complication should be noted
because no known lattice fermions have 
the full chiral symmetry on finite lattices\cite{nielsen81}.
Two conventional lattice fermions are the staggered and Wilson fermions.
In the formulation of staggered fermions \cite{susskind77},
the flavor-chiral symmetry 
${\rm SU}(N_F)_L\times{\rm SU}(N_F)_R\times{\rm U}(1)$
of massless $N_F$-flavor QCD %in the continuum limit
is explicitly broken down to 
${\rm U}(N_F/4)_L\times{\rm U}(N_F/4)_R$ at $a>0$. 
Moreover, this action is local only when $N_F$ is a multiple of 4. 
For the physically interesting cases $N_F=2$ and 3, a usual trick is 
to modify by hand the power of the fermionic determinant in the 
numerical path-integration. 
This necessarily makes the action non-local, 
that poses conceptually and technically difficult problems. 
With the Wilson fermion action \cite{wilson77}, on the other hand, 
the flavor symmetry is manifest also on the lattice, 
and the action is local for any $N_F$. 
This feature is quite attractive in a study of the $N_F$ dependence
of the QCD transition.
However, the chiral symmetry is explicitly broken at $a>0$, that 
requires additional tedious analyses of chiral properties. 

These broken symmetries of lattice fermions are 
expected to restore in the continuum limit. 
On a coarse lattice used in a finite temperature simulation, however, 
we may encounter sizable deviations from the scaling behavior
expected in the continuum limit.
Therefore, the appearance of the O(4) scaling is also a useful 
touchstone to test the recovery of the chiral symmetry on 
the lattice when the chiral transition is of second order.

\subsection{Results with staggered quarks}
\label{sec:JLQCD}

The O(4) scaling was first tested on the lattice for staggered quarks 
by the Bielefeld group\cite{KarschLaermann}. 
Based on simulations on an $8^3\times4$ lattice at $m_qa=0.02$,
0.0375, and 0.075 using the standard action, they obtained 
$z_g = 0.77(14)$, $z_m = 0.79(4)$, and $z_t = 0.65(7)$, 
where the corresponding O(4) values\cite{KanayaKaya} are 0.537(7), 
0.794(1), and 0.331(7). 
The result for $z_m$ is consistent with the O(4) value 
while other exponents are in disagreement with the O(4) values. 

\begin{figure}[t]
\centerline{
\epsfxsize=7cm \epsfbox{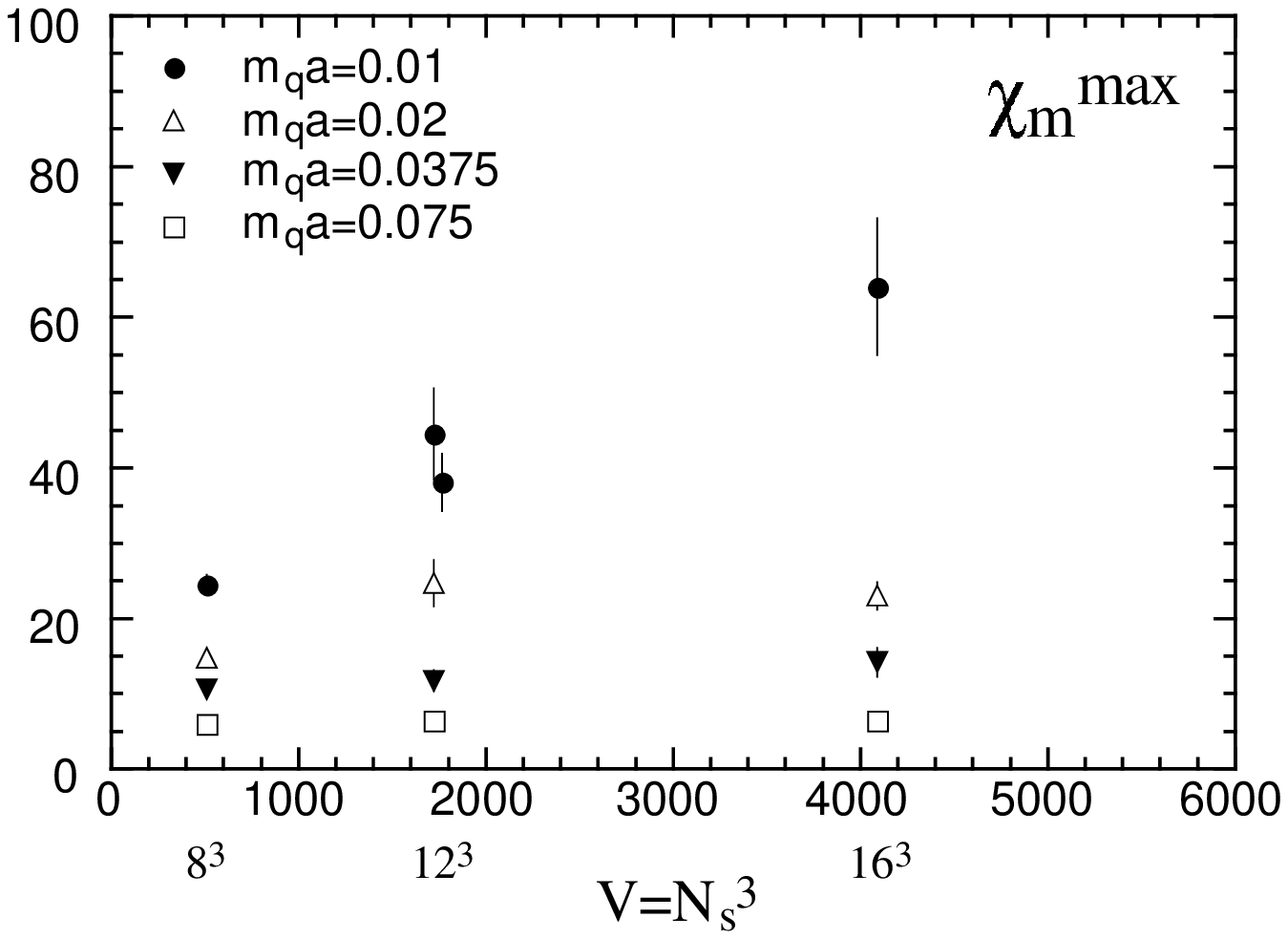}
\epsfxsize=7cm \epsfbox{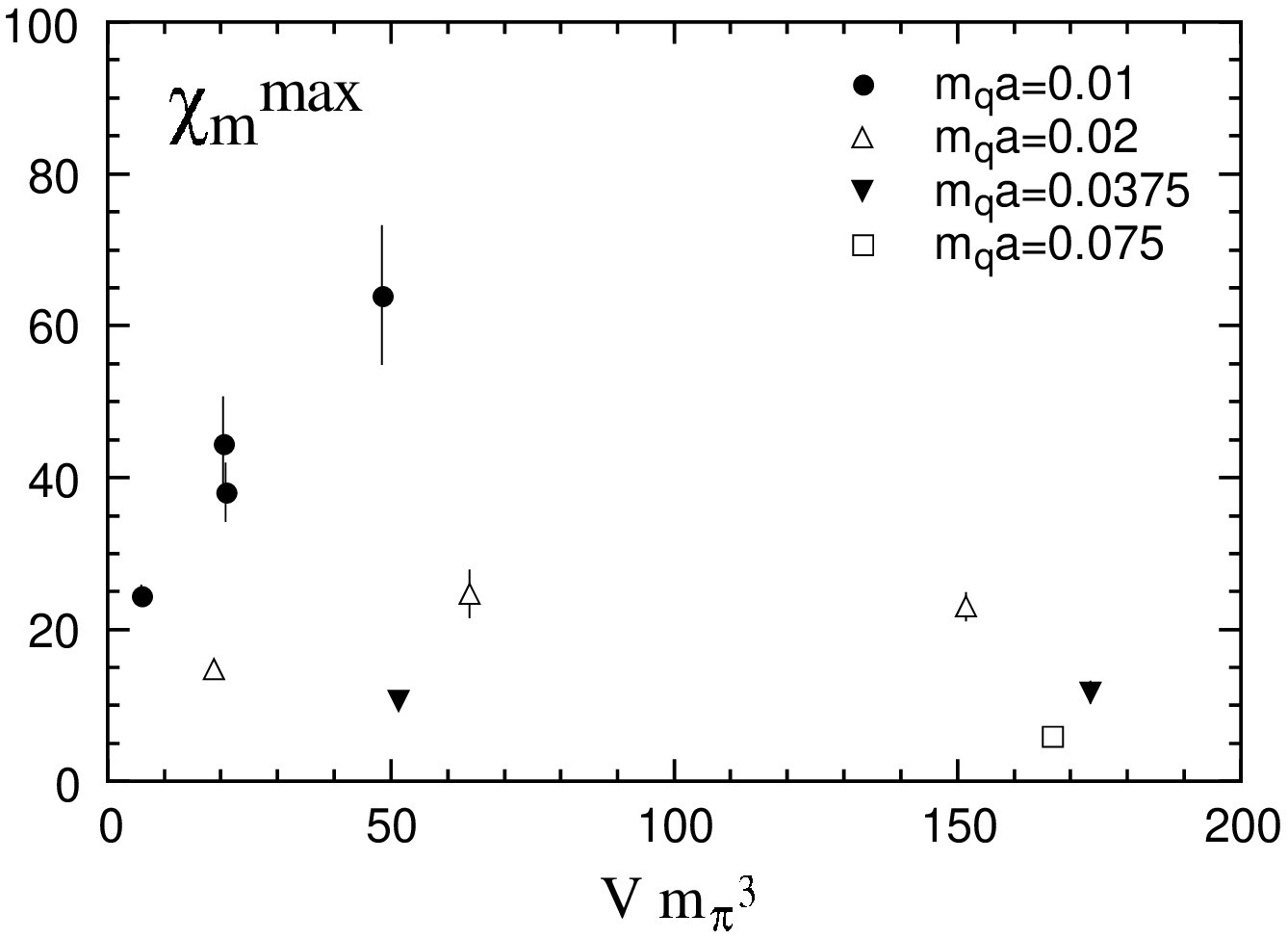}
}
\vspace{-0.2cm}
\caption{
(a) Peak height of the magnetic susceptibility 
for $N_F=2$ QCD with staggered quarks 
as a function of the spatial lattice volume $N_s^3$.
(b) The same data as a function of the lattice volume
rescaled by zero-temperature pion correlation length.
}
\label{fig:chim}
\end{figure}

Possible causes of the discrepancy are (i) $m_q$ is not small enough 
to see the critical behavior in the chiral limit, and (ii) 
the spatial lattice volume is not large enough to obtain the 
observables in the thermodynamic limit. 
Two additional caveats are in order for $N_F=2$ staggered quarks: 
(iii) The symmetry in the chiral limit at $a>0$ is O(2) 
instead of O(4). 
Practically, however, the values of the O(2) exponents are almost 
indistinguishable from the O(4) values with the present numerical accuracy.
(iv) The action is not local. 
Therefore, an assumption behind the universality argument 
can be violated so that some non-universal behavior may appear 
\cite{KanayaLat95}.
The correct continuum chiral limit with the O(4) symmetry will
be obtained only when we first take the continuum limit $a \rightarrow 0$ 
and then take the chiral limit. 
In addition to these points, 
we also have to check technical details in the numerical
simulation; the accuracy of the methods to simulate the 
system, such as the the finite step-size error and the dependence
on the convergence criterion for fermion matrix inversion.

A systematic study of the quark mass dependence as well as the lattice
volume dependence is in order.
We performed a series of simulations on
$8^3\times4$, $12^3\times4$, and $16^3\times4$ lattices at 
$m_qa=0.01$, 0.02, 0.0375, and 0.075 \cite{JLQCDfinT}.
The Bielefeld group also extended their study to larger spatial 
lattices \cite{KarschLaermann2}.
The results obtained are consistent with each other.
It turned out that determination of critical exponents on 
$8^3\times4$ lattices suffers from a sizable finite lattice-size 
effect for $m_qa < 0.0375$.

From the lattice-size dependence of the magnetic susceptibility $\chi_m$
for fixed value of quark mass, 
shown in Fig.~\ref{fig:chim}(a),
we conclude that the transition is a crossover for $m_qa\geq 0.02$; 
the peak height $\chi_m^{\rm max}$ for $m_qa=0.02$
stabilizes on spatial lattices larger than $12^3$.
For $m_qa=0.01$, on the other hand, $\chi_m^{\rm max}$ is increasing
up to our largest spatial lattice of $16^3$. 
If this increase is maintained up to infinite volume, then we have a 
first order transition at this quark mass. 
However, examining the lattice volume dependence of Monte Carlo time 
histories and histograms, we could not find any 
clear indication of a first-order transition at $m_qa=0.01$.
Furthermore,
when we rescale the lattice volume by zero-temperature pion correlation length,
we find that the lattice volume $16^3$
for $m_qa=0.01$ approximately
corresponds to the volume $12^3$ for $m_qa=0.02$,
where the increase of $\chi_m^{\rm max}$ terminates [Fig.~\ref{fig:chim}(b)].  
Therefore, it is possible that the increase of
$\chi_m^{\rm max}$ for $m_qa=0.01$ seen in our data is a
transient effect.

Assuming that the finite size effect is sufficiently small on the
$16^3\times4$ lattice, we fit the data at the four values of $m_qa$.
We find $z_g=0.64(5)$, $z_m=1.03(9)$, and $z_t=0.82(12)$.
(Removing the data for $m_qa=0.01$ gives slightly smaller
but consistent values with larger errors.)
The results for $z_g$ and $z_m$ sizably deviate 
from the O(2) or O(4) values.
We also studied other exponents related to energy expectation value, 
which also largely deviate from O(2) and O(4) values.
On the other hand,
the identity $z_g+z_m-z_t=1$ expected for a second-order fixed point with
two relevant operators is approximately satisfied.
Thus, 
the exponents are consistent with a second-order transition at $m_q=0$.
Our study of the scaling function also shows a similar trend: 
With measured values of exponents, data for the susceptibility $\chi_m$ 
exhibits a reasonable scaling as a function of $\beta$ and $m_qa$,
although results are much worse when we adopt the O(4) exponents.

In summary, we find the determination of the nature of the two-flavor
chiral transition with staggered quarks using the standard action 
to involve subtle problems.  
While our data so
far do not contradict a second-order transition at $m_q=0$, the exponents
take quite unexpected values, at least in the range $m_qa \geq 0.01$.  
Evidently further work, possibly on larger
spatial sizes and smaller quark masses, is needed to clarify this
important problem.

\subsection{Results with Wilson quarks}
\label{sec:QCDPAX}

We now study the issue using Wilson quarks.
It turned out that Wilson quarks in the standard action lead to 
several unexpected phenomena on lattices with $N_t=4$ and 6:
On these lattices, the transition becomes once very sharp when 
we increase $m_q$ from the chiral limit\cite{MILC46,ourStandard96},
in contrary to the expectation in the continuum limit that 
the chiral transition becomes weaker when we increase $m_q$.
Together with other strange behaviors of physical quantities 
near the transition point, this phenomenon is identified as an effect of 
lattice artifacts\cite{ourStandard96}. 
Therefore, we apply an improved action. 

\begin{figure}[t]
\begin{center}\leavevmode
\epsfxsize=13cm \epsfbox{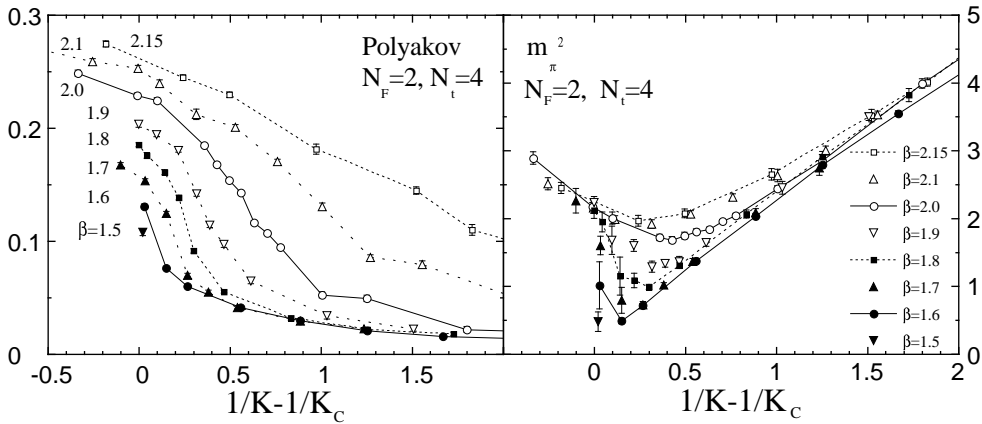}
\end{center}
\vspace{-0.2cm}
\caption{
The Polyakov loop and the pion screening mass 
obtained with Wilson quarks with a RG improved action
\protect\cite{ourPRL97}.
In these figures, the conventional notation $\beta \equiv 6/g^2$ 
is used, not to be confused with a critical exponent.
Larger $\beta$ corresponds to a higher temperature.
The horizontal axis $1/K-1/K_C$ is proportional to $m_q a$.
}
\label{fig:iF2T4PMpi}
\end{figure}

With the \RW\ action discussed in Sec.~\ref{sec:CPPACS}, we first 
find that the lattice artifacts observed with the standard action
are well removed\cite{ourPRL97,ImFQcdTsukuba}. 
We also find that the physical quantities are quite smooth around the
transition point at $m_q>0$, as shown in Fig.~\ref{fig:iF2T4PMpi}.
The straight line envelop of $m_\pi^2$ at finite temperature ($N_t=4$) 
shown in Fig.~\ref{fig:iF2T4PMpi}(b) 
agrees with $m_\pi^2$ obtained at low temperature ($N_t=8$), 
and corresponds to the PCAC relation $m_\pi^2 \propto m_q$ expected in the 
low-temperature phase.
The smoothness of the physical observables strongly suggests that 
the transition is a crossover at $m_q > 0$. 

Concerning the nature of the transition in the chiral limit,
we find that 
the transition becomes monotonically weaker when we increase $6/g^2$
(see Fig.~\ref{fig:iF2T4PMpi}).
Because the transition point shifts to larger $6/g^2$ at larger $m_q$, 
increasing $6/g^2$ corresponds to increasing $m_q$ for the transition.
We also note that $m_\pi^2$ in the chiral limit monotonically
decreases to zero as we decrease temperature from above 
towards the chiral transition point. % $\beta_{ct} \simeq 1.35$, 
$m_\pi^2$ at the transition temperature for finite $m_q$ 
also shows a similar monotonic decrease when we decrease $m_q$.
These results suggest that the chiral transition is continuous. 

\begin{figure}[t]
\centerline{
\epsfxsize=6.5cm \epsfbox{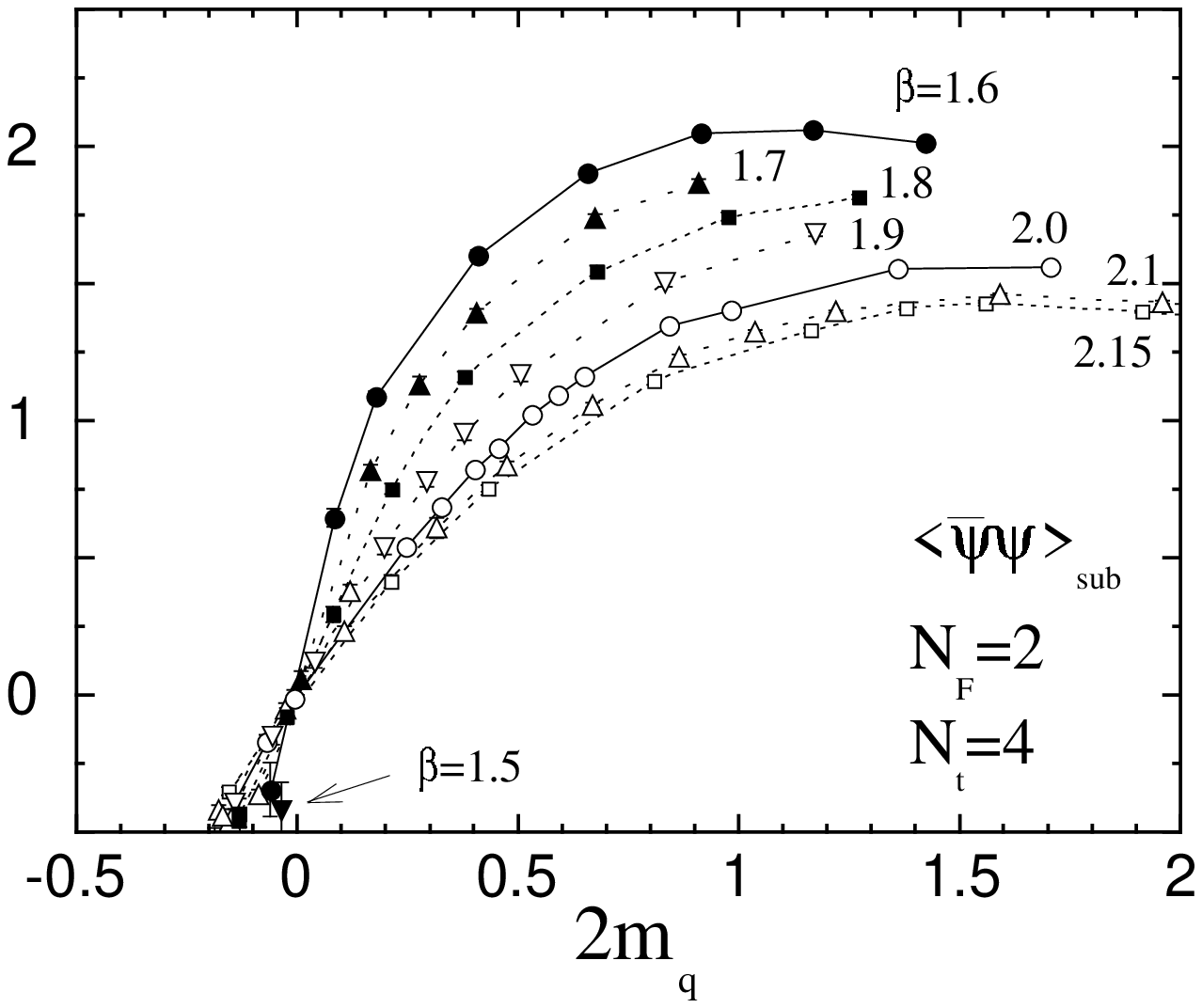}
\makebox[2mm]{}
\epsfxsize=6.7cm \epsfbox{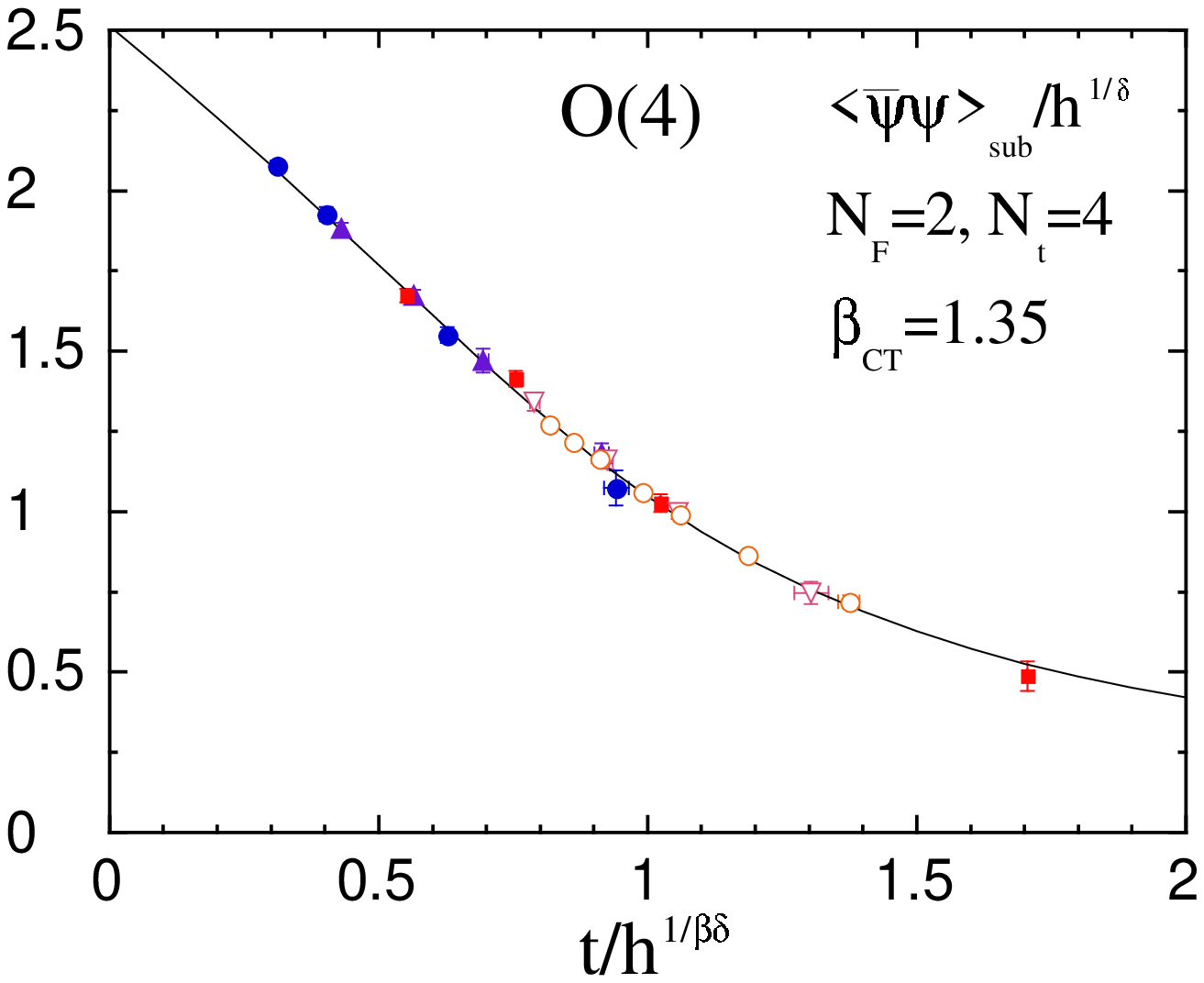}
}
\vspace{-0.2cm}
\caption{(a) 
Chiral condensate as a function of $h=2m_qa$ 
for Wilson quarks with a RG improved action\protect\cite{ourPRL97}.
(b)
Best fit for the scaling function with O(4) exponents.
The plot contains all data of (a) 
within the range $0 < 2m_q a < 0.8$ and $6/g^2 \leq 2.0$. 
Solid curve represents the scaling function obtained in an O(4) 
spin model. 
$\beta_{CT}$ is the value of $6/g^2$ at the chiral transition point.
In these figures, the conventional notation $\beta \equiv 6/g^2$ is used,
not to be confused with the critical exponent $\beta$ 
appearing in the combination $t/h^{1/\beta\delta}$.
}
\label{fig:pbp}
\end{figure}

For a more decisive test about the nature of the transition, 
a scaling study is required.
From the universality argument 
we expect that magnetization $M$ near the second order 
transition point can be described by a single scaling function: 
\begin{equation}
M / h^{1/\delta} = f(t/h^{1/\beta\delta}),
\end{equation}
where $h$ is the external magnetic field 
and $t=[T-T_c]/T_c$ the reduced temperature. 
When the QCD transition is of second order in the chiral limit, 
the chiral condensate 
should satisfy this scaling relation %(\ref{eq:universality}) 
with O(4) exponents 
$1/\beta\delta = 0.537(7)$ and $1/\delta = 0.2061(9)$ 
\cite{KanayaKaya} 
and the O(4) scaling function $f(x)$.
Our results for $M$ are shown in Fig.~\ref{fig:pbp}(a). 
We make a fit of $M$ to the scaling function
obtained for an O(4) model \cite{Toussaint},
by adjusting $\beta_{ct}$ and the scales for $t$ and $h$, 
with the exponents fixed to the O(4) values.
Figure \ref{fig:pbp}(b) shows our result
with $\chi^2/df = 0.61$. 
The scaling ansatz works remarkably well with the O(4) exponents. 
Our recent study shows that the situation holds also when we include data
in the $t \leq 0$ region\cite{ImFQcdTsukuba}.
On the other hand, a change of the exponents quickly makes the fit worse:
For example, when we use the MF exponents suggested by Koci\'c and
Kogut as a possibility for two-flavor QCD\cite{KocicKogut},
the data no more falls on the MF scaling function. 

The success of this scaling test with the O(4) exponents 
strongly suggests that the chiral transition is of second order 
in the continuum limit. 
It also indicates that 
the chiral violation due to the Wilson fermion action is sufficiently 
small with our improved action, 
for the values of $m_q$ and $6/g^2$ studied here. 

\section{Summary}

A simulation of finite temperature QCD sometimes suffers from sizable
lattice artifacts caused by the coarseness of the lattice. 
Due to a requirement of large spatial lattice size for the 
thermodynamic limit, and also due to 
physical requirements to study a wide range of the parameter space, 
it is hard to remove these lattice artifacts when we use the 
standard lattice actions.
Recent developments in improved lattice actions opened us a
possibility to perform accurate simulations of finite temperature QCD
with the present power of computers.
A comparative study of improved full QCD actions shows that an efficient
reduction of lattice artifacts can be achieved already on lattices 
with $a \sim 0.2$fm, as used in major finite temperature simulations.

Among the topics of finite temperature QCD, 
intensive lattice studies have been made recently on 
the nature of the chiral transition in QCD with two degenerate light quarks. 
Understanding it is an important step toward the clarification of 
the transition in the real world including the non-degenerate s quark.
Numerical simulations show that the transition for $N_F=2$ is 
analytic crossover down to a very small value of the quark mass. 
In order to further determine the nature of the transition in 
the chiral limit, we have to study the scaling property and compare 
it with the O(4) scaling expected theoretically 
for the case of a second order transition.

Using the standard action for the staggered quark, 
we find the determination of the nature of the two-flavor
chiral transition to involve subtle problems.  While our data 
do not contradict a second-order transition at $m_q=0$, 
the critical exponents take quite unexpected values. 
It is not yet clear whether this strange behavior is caused by 
large sub-leading terms in the scaling relations, 
or by the non-universality 
caused by the non-locality of the two-flavor staggered quark action. 
If the latter is the case, application of an improved action may
solve the problem.  

A study of the issue using the Wilson quark has been difficult 
due to severe lattice artifacts encountered 
when we use the standard action.
We find that these lattice artifacts are removed 
when we use an improved gauge action combined with the standard Wilson 
quark action. 
Furthermore, we find that the chiral condensate remarkably follows the scaling 
relation with O(4) exponents and O(4) scaling function. 
This strongly suggests that the chiral transition in two-flavor QCD is 
of second order. 

This second order transition for the case $m_s=\infty$ should turn 
into a first order transition when we decrease $m_s$, as 
observed for the degenerate $N_F=3$ case.
Because $m_s \simeq 150$ -- 200 MeV is just of the same order 
of magnitude as the expected values of $T_c$, 
we have to fine-tune the value of $m_s$ 
in order to study the nature of the transition in the real world. 
Unfortunately, the lattice simulations performed so far for the 
$N_F=3$ and $N_F=2+1$ cases\cite{Columbia,ourStandard96} 
do not have that accuracy yet 
because, using standard lattice actions, we have sizable 
systematic errors due to lattice artifacts.
At the moment, both possibilities of a first order transition 
and an analytic crossover are remaining for the physical point.

Encouraged with the success of the scaling study for two-flavor QCD 
using an improved action, we began test simulations for $N_F=3$ and 
$2+1$ applying the same improved action\cite{ImFQcdTsukuba}.
Many lattice groups began to simulate finite temperature QCD 
using improved actions.
It may, however, take several more years to 
re-accumulate the basic data in a wide range of the many-dimensional 
full QCD parameter space, that is required to clarify the nature 
of the QCD transition in the real world.

\vspace{3mm}
This work is in part supported by the Grants-in-Aid
of Ministry of Education (Nos. 08NP0101, 09304029),
and also in part by the Supercomputer Project (No.97-15) 
of High Energy Accelerator Research Organization (KEK).

\end{document}